\documentclass{iaus}
\usepackage{graphicx}

\def \littleprime{\ifmmode{\scriptscriptstyle \prime}\else{\hbox{$\scriptscriptstyle \prime$ }}\fi} 
\def \arcsec{\raise .9ex \hbox{\littleprime\hskip-3pt\littleprime}} 

\def\Vc{V_{\rm c}}  
\def\cs{\sigma_0} 
\def\C28{\rm C_{28}} 

\title[The $\Vc-\cs$ Relation of Galaxies] 
{The $\Vc-\cs$ Relation of Galaxies}

\author[Courteau et~al.] 
{St\'ephane Courteau, Michael McDonald, and Lawrence M. Widrow}
\affiliation{Department of Physics, Engineering Physics \& Astronomy, Queen's 
  University, Kingston, Ontario, Canada. 
\break email: courteau,mcdonald,widrow@astro.queensu.ca}

\pubyear{2008}
\volume{245}
\pagerange{001--005}
\date{31 August 2007}
\jname{Formation and Evolution of Galaxy Bulges}
\editors{M. Bureau, E. Athanassoula \& B. Barbuy, eds.}
\begin{document}

\maketitle

\begin{abstract}
Courteau \etal\ (2007a) reported on the dependence of the ratio of a galaxy's 
maximum circular velocity, $\Vc$, to its central velocity dispersion, $\cs$, 
on morphology, or equivalently total light concentration. 
This $\Vc-\cs-$concentration relation, which involves details about the local 
and global galaxy physics, poses a fundamental challenge for galaxy structure
models.  Furthermore, not only must these models reproduce the $\Vc-\cs$ relation 
and its various dependences, they must simultaneously match other fundamental 
scaling relations such as the velocity-size-luminosity and color-luminosity 
relations of galaxies.  We focus here on the interpretation of parameters that
enter the $\Vc-\cs$ relation to enable proper data-model comparisons and follow-up 
studies by galaxy modelers and observers. 
\keywords{galaxies: fundamental parameters, galaxies: kinematics and dynamics}
\end{abstract}

\firstsection 
\firstsection 
\bigskip

\section{Introduction}

The $\Vc - \cs$ relation is of great interest for galaxy structure 
models since it links two quantities that depend separately on global 
and local physics.  The galaxy circular velocity, $\Vc=\sqrt{GM(r)/r}$
where $M(r)$ is the total mass within $r$ of the center, is directly 
related to the total mass of the galaxy whereas the central velocity 
dispersion, $\cs$, is a measure of the local central potential.  These 
two quantities could in principle be independent.  It is well known 
that the brightest, bulge dominated (E, S0, and some Sa), galaxies 
obey closely the relation $\Vc = \sqrt{2} \cs$ expected for isothermal 
gravitational systems (Whitmore \etal\ 1979; Courteau \etal\ 2007a; Ho 2007).  
The lower surface brightness regime, dominated by later-type spiral and dwarf 
galaxies, however departs from the isothermal solution and the ratio 
$\Vc/\cs$ here scales with surface brightness (Pizzella \etal\ 2005; 
Buyle \etal\ 2006; Courteau \etal\ 2007a; Ho 2007) or, equivalently, 
total light concentration or morphological type.  The correlation between 
concentration index and Hubble type is discussed by Strateva \etal\
(2001) in the context of SDSS galaxies.  Figure 1 encapsulates 
the dependence of the $\Vc-\cs$ relation on the galaxy Hubble type. 
The current data (see Courteau \etal\ 2007a for details) show
that $\Vc \approx 5 \cs$ for dIrr galaxies suggesting that these
galaxies live in very dominant dark matter halos. 

\begin{figure}
 \includegraphics[width=1.00\textwidth]{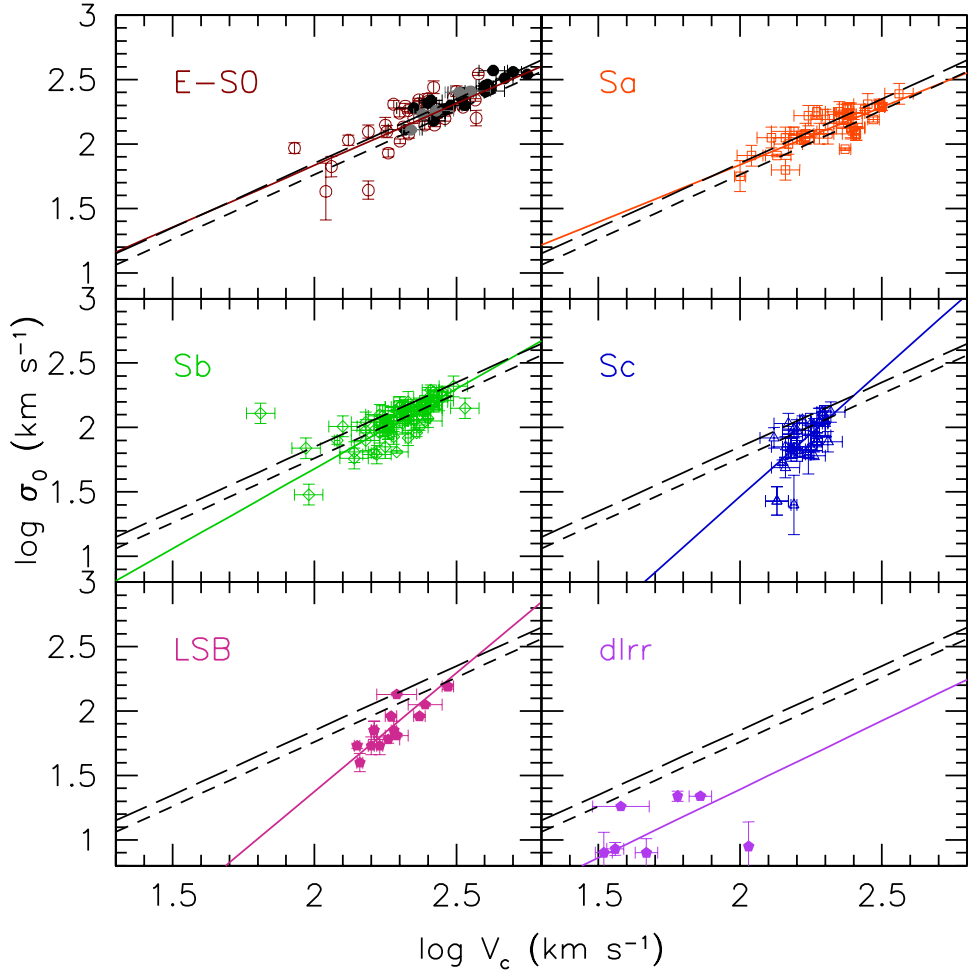}
  \caption{$\Vc - \cs$ relation for galaxies of different Hubbles types. 
   The long and short dashed lines represent the $\Vc = \sqrt{2} \cs$
   and $\Vc = \sqrt{3} \cs$ solutions respectively.  The solid and open
   circles in the upper left panel represent E and S0 galaxies, respectively.
   Galaxies with lower surface brightnesses depart progressively from the 
   nominal $\Vc = \sqrt{2} \cs$ solution for isothermal systems, as gauged
   by the Sc and LSB galaxies.  Taken at face value, the $\Vc - \cs$ 
   measures for dIrr galaxies suggest that these galaxies live in 
   very dominant dark matter halos. 
}

\label{fig:Vc6}
\end{figure}

In order to understand and, ideally, model the values of $\Vc$ and
$\cs$ and their dependence on each other, we must pay close attention
to their precise definition and how they are being measured.  

\section{Data}\label{sec:data}

Measurements of the circular velocity $\Vc$ differ for gas-rich and 
gas-poor systems.  For the former, $\Vc$ is often estimated to be 
the maximum deprojected orbital velocity, $V_{max}$, usually measured from
emission lines.  This is marginally correct so long as $r(V_{max})$ 
lies beyond the radius where non-circular velocities from asymmetric 
drift are dominant; for bright spiral galaxies, that radius can be 
as small as two disk scale lengths (e.g. Courteau \etal\ 2003; 
Spekkens \& Sellwood 2007) but it rises to more than three disk
scale lengths at lower surface brightnesses (Rhee \etal\ 2004; 
Valenzuela \etal\ 2007).  Neglect of this effect can have nefarious 
consequences on estimates of $\Vc$.  The measurement of $\Vc$ for 
gas-poor (spheroidal) galaxies  
is inferred via non-parametric dynamical modeling of the absorption line 
features and surface brightness profiles of these galaxies 
(Gerhard \etal\ 1998; Kronawitter \etal\ 2000; Gerhard \etal\ 2001; 
Cappellari \etal\ 2006).

The one-dimensional line-of-sight (projected) central velocity dispersion, 
$\cs$, is measured from selected absorption lines (e.g. Lick indices) 
typically within the central $1\arcsec$ or the radius corresponding to 
the slit width.  
A theorist would estimate $\cs$ first by building a model for the 
phase space distribution function (DF) of the stars and then 
integrating over the DF (see Binney \& Tremaine 1987, Eq. 4-57). 
Alternatively, one can generate an N-body representation of the 
galaxy from the DF and calculate the rms velocity along a given 
line of sight within a ``beam'' whose width is chosen to match 
that of the observations.  This N-body representation provides 
a suitable starting point for numerical simulations to study the 
formation of bars and bulges which in turn can affect $\cs$.  

Various observers correct $\cs$ to a standard aperture of $r_{e}/8$, 
where $r_e$ is the effective radius of the galaxy 
spheroid (e.g. Jorgensen \etal\ 1995; Cappellari \etal\ 2006; 
MacArthur \etal\ 2007).  
This definition is however awkward for disk dominated systems 
with a bulge as $r_e$ cannot be uniquely determined.  Fortunately, 
aperture effects may be small (Pizzella \etal\ 2004) and the 
correction to $\cs$ for aperture size is here neglected.  $\cs$ is 
therefore measured the same way for high and low surface brightness
galaxies.  Two other effects may bias the determination of $\cs$: 
net rotation of the bulge and contamination from the (largely rotating) 
disk.  The latter has been verified to be small, even in late-type 
spirals (Pizzella \etal\ 2004).  The rotation of the 
bulge can be estimated on a case-by-case basis and while most studies thus 
far have found little angular momentum in galaxy bulges more research on 
this topic, with the largest possible telescopes or numerical simulations, 
is called for.  Thus, much like Gebhardt \etal\ (2000), our aperture 
dispersions include a contribution 
from rotationally supported material (i.e., the rms velocity is measured 
relative to the systemic velocity, not relative to the local mean velocity).  
The measured dispersion also depends on the inclination of the galaxy; 
this can be estimated in principle by the strength of the stellar rotation.

$\Vc$ and $\cs$ are linked, as in Courteau \etal\ (2007a) and Ho (2007), 
by concentration (figures not shown here).  Our data base includes 
the concentration measure 
$C_{28}=5\log(r_{80}/r_{20})$ where $r_{80}$ and $r_{20}$ are the 
radii measured at 80\% and 20\% of total light.  Theory suggests 
that $\Vc/\cs$ is controlled by the degree of compactness of a galaxy 
as measured either by morphology, B/T, surface brightness, concentration, 
etc. (Courteau \etal\ 2007a; Pe\~narrubia \etal\ 2007).  A significant 
advantage of concentration is its non-parametric definition, independence
of photometric calibration, and weak dependence on inclination.  
Concentration indices however depend on the photometric bands and
our data base uses SDSS $i$-band images.  

It should be noted that Ho (2007) recently used integrated 21cm line widths
and central velocity dispersions from HyperLeda and SDSS for 792 galaxies 
spanning a broad range of Hubble types to reproduce the results presented
in Courteau \etal\ (2007a).  His concentrations used a ratio of SDSS 
$i$-band Petrosian radii enclosing 90\% and 50\% of the light.
There is global agreement between the two concentration
measures of Courteau \etal\ (2007a) and Ho (2007) albeit 
with noticeable scatter.  In spite of these concerns and 
differences in the global samples and data products, the 
complimentary studies of Courteau \etal\ (2007a) and Ho (2007)
come to close agreement about the dependence of the ratio 
$\Vc/\cs$ on concentration. 

Nonetheless, in order to reproduce Fig.~1 and the overall $\Vc-\cs-$concentration 
relation of galaxies, especially from theoretical/numerical stand-points
(which we have no room to discuss here but see Courteau \etal\ 2007a for 
an introduction), care must be taken to use quantities as defined above.  
Our complete data compilation is available at 
www.astro.queensu.ca/$\sim$courteau/data/VSigmaC28.dat. 

\section{Further developments}
Models of galaxy structure ought to reproduce the $\Vc-\cs$ relation 
and its dependence on concentration, in addition to matching other 
basic scaling relations of galaxies such as the velocity-size-luminosity 
relations reported in, for instance, Gnedin \etal\ (2006), Courteau \etal\ 2007b 
and Dutton \etal\ (2007).   Matching these relations at different wavebands, 
and thus accounting for the color-luminosity relation of galaxies, is another 
formidable challenge. 

One must bear in mind that the data reported here, as well as in Ho (2007), 
come from widely heterogeneous data bases.  While homogenization has been 
optimised, there is still no substitute to perfectly homogeneous data by 
design and we urge the community to invest in long-term dedicated surveys 
of kinematic parameters for galaxies of all types.  Only with well-understood 
dynamical measurements can we construct a complete and robust picture of
dynamical evolution of galaxies ranging from the central supermassive black 
holes and nuclear star clusters to the largest halo structures in galaxies. 

\bigskip 

\begin{acknowledgments}
We are grateful to Alessandro Pizzella, Anatoly Klypin, Sandra 
Faber and Andy Burkert for useful discussions that prompted various 
comments reported here.  S.C. and L.M.W. acknowledge the support of 
NSERC through respective Discovery grants.   Thank you Martin for 
a superb symposium!

\end{acknowledgments}

\newpage 

\begin{discussion}

\discuss{Emsellem}{Are the models by Widrow \& Dubinski (2005 [WD05]) 
reported in Courteau \etal\ (2007a) and shown in your presentation [but 
omitted in this paper] using values for $\Vc$ on $\cs$ consistent with 
observed values?}

\discuss{Courteau}{To all possible extent, yes.  The GalactICS model
(Kuijken \& Dubinski 1995; WD05) provides self-consistent equilibrium 
phase-space distribution functions for bulge-disk-halo systems.  
So, in principle, it is possible 
to compute $\cs$ the way observers do as a mix of all bulge and disk 
particles within some fixed apertures
projected on the sky.   
The development of such models is a life-long affair though and better 
data-model agreements should be expected with improved treatments of 
density profile cores, presence of spiral arms, hydrodynamical effects, 
etc.  I was delighted to learn at this conference that Herv\'e Wozniak 
is already attempting to reproduce the $\Vc-\cs-\C28$ relation.  Unlike 
WD05, his models include gas dynamics though the significance of this 
has yet to be fleshed out.}

\discuss{Bendo}{Rather than using concentration which depends on inclination, 
why not use a more robust parameter for morphology such as M$_{20}$ as 
defined by Lotz et al. 2004?}

\discuss{Courteau}{The motivation of this exercise is to use any 
unbiased morphological indicator that can be linked to a theoretical 
interpretation of the dependence of $\Vc$ on $\cs$.  Hubble type, surface 
brightness, concentration, Gini, M$_{20}$, and the likes are all useful.  
However, morphological types usually have an observer bias, surface
brightnesses depend on the absolute magnitude calibration, etc.  
We chose concentration because it is a straightforward parameter that 
can be computed with unscaled profiles ($C_{28}$ does depend on waveband 
though) for observers and theorists alike.  Note also that this 
concentration parameter is only weakly inclination dependent since 
the similar inclination dependences of the radii $r_{80}$ and 
$r_{20}$ essentially cancel out [a figure was shown to support 
this argument].  The inclination dependence is even weaker for 
the concentration parameter of Ho (2007).}

\discuss{Peletier}{Could you use a parameter from pseudo-bulges 
to disentangle the $\Vc-\cs$ relation?}

\discuss{Courteau}{Graham \etal\ (2001) have already discussed 
the correlation of the bulge light concentration, in lieu of 
the central velocity dispersion, with the central supermassive
black hole mass.  Because we are interested in matching local
($\cs$) with global ($\Vc$) physics in galaxies, the specifics 
of any putative pseudo-bulge are of lesser interest to us at 
the moment.  This is not to say that the study of the formation 
and evolution of bulges of all forms isn't fascinating in its 
own right!}

\end{discussion}

\end{document}